\documentstyle[12pt]{article}

\begin{document}

\title{Black Holes and Thermodynamics}
\author{Robert M. Wald\\
         {\it Enrico Fermi Institute and Department of Physics}\\
         {\it University of Chicago}\\
         {\it 5640 S. Ellis Avenue}\\
         {\it Chicago, Illinois 60637-1433}}
\maketitle

\begin{abstract}

We review the remarkable relationship between the laws of black hole
mechanics and the ordinary laws of thermodynamics. It is emphasized
that -- in analogy with the laws of thermodynamics -- the validity the
laws of black hole mechanics does not appear to depend upon the
details of the underlying dynamical theory (i.e., upon the particular
field equations of general relativity). It also is emphasized that a
number of unresolved issues arise in ``ordinary thermodynamics'' in
the context of general relativity. Thus, a deeper understanding of the
relationship between black holes and thermodynamics may provide us
with an opportunity not only to gain a better understanding of the
nature of black holes in quantum gravity, but also to better
understand some aspects of the fundamental nature of thermodynamics
itself.

\end{abstract}
\newpage

\section{Introduction}

Undoubtedly, one of the most remarkable developments in theoretical
physics to have occurred during the past twenty five years was the
discovery of a close relationship between certain laws of black hole
physics and the ordinary laws of thermodynamics. It appears that these
laws of ``black hole mechanics'' and the laws of thermodynamics are two
major pieces of a puzzle that fit together so perfectly that there can
be little doubt that this ``fit'' is of deep significance. The
existence of this close relationship between these laws may provide us
with a key to our understanding of the fundamental nature of
black holes in a quantum theory of gravity, as well as to our
understanding of some aspects of the nature of thermodynamics
itself. The aim of this article is to review the nature of the
relationship between the black hole and thermodynamics laws. Although
some notable progress has been made, many mysteries remain.

It was first pointed out by Bekenstein \cite{b1} that a close
relationship might exist between certain laws satisfied by black holes
in classical general relativity and the ordinary laws of
thermodynamics. The area theorem of classical general relativity
\cite{h} states that the area, $A$, of a black hole can never decrease
in any process
\begin{equation}
\Delta A \geq 0.
\label{bh2}
\end{equation}
Bekenstein noted that this result is closely analogous to the
statement of ordinary second law of thermodynamics: The total entropy,
$S$, of a closed system never decreases in any process
\begin{equation}
\Delta S \geq 0.
\label{t2}
\end{equation}
Indeed, Bekenstein proposed that the area of a black hole (times a
constant of order unity in Planck units) should be interpreted as its
physical entropy.

A short time later, the analogy between certain laws of black hole
physics in classical general relativity and the laws of thermodynamics
was developed systematically by Bardeen, Carter, and Hawking
\cite{bch}. They proved that in general relativity, the surface
gravity, $\kappa$, of a stationary black hole (defined by
eq.(\ref{kappa}) below) must be constant over the event horizon of the
black hole. They noted that this result is analogous to the zeroth law
of thermodynamics, which states that the temperature, $T$, must be
uniform over a body in thermal equilibrium. Finally, Bardeen, Carter,
and Hawking proved the ``first law of black hole mechanics''. In the
vacuum case, this law states that the differences in mass, $M$, area,
$A$, and angular momentum, $J$ of two nearby stationary black holes
must be related by
\begin{equation}
\delta M = \frac{1}{8 \pi} \kappa \delta A + \Omega \delta J,
\label{bh1}
\end{equation}
where $\Omega$ denotes the angular velocity of the event horizon
(defined by eq.(\ref{xi}) below). Additional terms may appear on the
right side of eq.(\ref{bh1}) when matter fields are present. They
noted that this law is closely analogous to the ordinary first law of
thermodynamics, which states that differences in energy, $E$, entropy,
and other state parameters of two nearby thermal equilibrium states of
a system is given by
\begin{equation}
\delta E = T \delta S + {\mbox{\rm ``work terms''}}.
\label{t1}
\end{equation}

If we compare the zeroth, first, and second laws of ordinary
thermodynamics with the corresponding laws of black hole mechanics, we
see that the analogous quantities are, respectively, $E
\longleftrightarrow M$, $T \longleftrightarrow \alpha \kappa$, and $S
\longleftrightarrow A/8 \pi \alpha$, where $\alpha$ is a undetermined
constant. Even in the context of classical general relativity, a hint
that this relationship might be of physical significance arises from
the fact that $E$ and $M$ represent the same physical quantity, namely
the total energy of the system. However, in classical general
relativity, the physical temperature of a black hole is absolute zero,
so there can be no physical relationship between $T$ and
$\kappa$. Consequently, it also would be inconsistent to assume a
physical relationship between $S$ and $A$. For this reason, at the
time the paper of Bardeen, Carter, and Hawking appeared, most
researchers (with the notable exception of Bekenstein) viewed the
analogy between the black hole and thermodynamical laws as a
mathematical curiosity, devoid of any physical significance.

That view changed dramatically with Hawking's discovery \cite{h2}
that, due to quantum particle creation effects, a black hole radiates
to infinity all species of particles with a perfect black body
spectrum, at temperature (in units with $G=c=\hbar=k=1$)
\begin{equation}
T = \frac{\kappa}{2 \pi} .
\label{T}
\end{equation}
Thus, $\kappa/2 \pi$ truly is the {\em physical} temperature of a
black hole, not merely a quantity playing a role mathematically
analogous to temperature in the laws of black hole mechanics. This
left little doubt that the laws of black hole mechanics must
correspond physically to the laws of thermodynamics applied to a
system consisting of a black hole. As will be discussed further below,
it also left little doubt that $A/4$ must represent the physical
entropy of a black hole in general relativity.

Thus, Hawking's calculation of particle creation effectively gave a
resoundingly positive answer to the question of whether there exists
any physical significance to mathematical the relationship between the
laws of black hole mechanics and the laws of thermodynamics. This
conclusion is particularly intriguing, since, as I shall review in
sections II and III, the the derivations of the laws of black hole
mechanics are so different in nature from those of the laws of
thermodynamics that it is hard to see how it is possible that these
laws could really be ``the same''. As I will discuss further in
section IV, this conclusion also raises a number of new questions and
issues, most prominently that of providing a physical explanation for
the origin of black hole entropy.

\section{The nature of the laws of ordinary thermodynamics}

The ordinary laws of thermodynamics are not believed to be fundamental
laws in their own right, but rather to be laws which arise from the
fundamental ``microscopic dynamics'' of a sufficiently complicated
system when one passes to a macroscopic description of it. The great
power and utility of the laws of thermodynamics stems mainly from the
fact that the basic form of the laws does not depend upon the details
of the underlying ``microscopic dynamics'' of particular systems and,
thus, the laws have a ``universal'' validity -- at least for a very
wide class of systems.

The analysis of how the laws of ordinary thermodynamics arise applies
to a classical or quantum system with a large number of degrees of
freedom, whose time evolution is governed by Hamiltonian dynamics. It
is important to emphasize that it is crucial here, at the outset, that
there be a well defined notion of ``time translations'' (to which the
Hamiltonian is conjugate), and that the Hamiltonian, $H$, (and, thus,
the dynamics) be invariant under these time translations. It then
follows that the total energy, $E$, of the system (i.e., the value of
$H$) is conserved.

We now shall focus on the case of a classical dynamical system. By the
previous remark, it follows that each dynamical orbit in phase space
is confined to its ``energy shell'' $\Sigma_E$, i.e., the hypersurface
in phase space defined by the equation $H(x) = E = {\rm
constant}$. The crucial assumption needed for the applicability of
thermodynamical laws to such a system is that ``generic'' orbits in
phase space behave ``ergodically'' in the sense that they come
arbitrarily close to all points of $\Sigma_E$, spending ``equal times
in equal volumes''; equivalently, the total energy of the system is
the only nontrivial constant of motion for generic orbits. (By a
slight modification of these arguments, the laws also can accomodate
the presence of a small number (compared with the number of degrees of
freedom) of additional constants of motion -- such as the angular
momentum of a rotationally invariant system.) Of course, even when
such ergodic behavior occurs, it would take a dynamical orbit an
infinite amount of time to completely ``sample'' $\Sigma_E$. The
degree of ``sampling'' which is actually needed for the applicability
of thermodynamics depends upon what observable (or collection of
observables) of the dynamical system is being measured -- or, in more
common terminology, the amount of ``coarse graining'' of phase space
that one does. For a ``fine-grained'' observable (corresponding to
measuring detailed information about the ``microscopic degrees of
freedom'' of the system), the ``sampling'' of the energy shell must be
extremely good, and the amount of time needed for this ``sampling''
will be correspondingly long, thereby making the laws of
thermodynamics inapplicable or irrelevant for the system. However, for
the types of ``macroscopic, coarse-grained'' observables, ${\cal O}$,
usually considered for systems with a huge number of degrees of
freedom, the ``sampling'' need only be quite modest and the required
``sampling time'' (= the timescale for the system to ``reach thermal
equilibrium'') is correspondingly short. One may then get considerable
predictive power from the laws of thermodynamics about the values of
${\cal O}$ which one would expect to observe for the system under
various physical conditions.

The {\em statistical entropy}, ${\cal S}_{\cal O}$, of a classical
dynamical system relative to a macroscopic, coarse-grained observable
(or collection of observables), ${\cal O}$, is defined to be the
observable whose value at point $x$ on $\Sigma_E$ is the logarithm of
the volume of the region of the energy shell at which ${\cal O}$ takes
the same value as it does at $x$, i.e.,
\begin{equation}
{\cal S}_{\cal O}(x) = \ln [{\rm vol} ({\cal R}_x)]
\label{calS}
\end{equation}
where
\begin{equation}
{\cal R}_x = \{y \in \Sigma_E | {\cal O}(y) = {\cal O}(x)\}.
\label{calR}
\end{equation}
For the types of coarse-grained observables ${\cal O}$ which are
normally considered, the largest region, ${\cal R}^{\rm max}$, of the
form (\ref{calR}) will have a volume nearly equal to that of the
entire energy shell $\Sigma_E$.  If dynamical orbits sample $\Sigma_E$
and spend ``equal times in equal volumes'', then we would expect
${\cal S}_{\cal O}$ to increase in value until it reaches its maximum
possible value, namely, $\ln [{\rm vol} ({\cal R}^{\rm max})] \simeq
\ln [{\rm vol} (\Sigma_E)]$.  Subsequently, ${\cal S}_{\cal O}$ should
remain at that value for an extremely long time. During this extremely
long period, the value of ${\cal O}$ remains unchanged, so no change
would be perceived in the system, and the system would be said to have
achieved ``thermal equilibrium''.

The {\em thermodynamic entropy}, $S$, of the system is defined by
\begin{equation}
S = \ln [{\rm vol} (\Sigma_E)]
\label{S}
\end{equation}
Unlike ${\cal S}_{\cal O}$, $S$ is not an observable on phase space
but rather a function on a low dimensional {\em thermodynamic state
space} comprised by the total energy, $E$, of the system, and any
parameters (such as, for example, an external magnetic field)
appearing in the Hamiltonian which one might contemplate varying,
together with any additional constants of motion for the system (such
as, the total angular momentum for a rotationally invariant
system). These variables characterizing thermodynamic state space are
usually referred to as {\em state parameters}. The temperature, $T$,
is defined by
\begin{equation}
\frac{1}{T} = \frac{\partial S}{\partial E} ,
\label{temp}
\end{equation}
where the remaining state parameters are held fixed in taking this
partial derivative. Like $S$, the temperature, $T$, is a function on
thermodynamic state space, not an observable on phase space.

It follows from the above discussion that when a system is in thermal
equilibrium, its statistical entropy, ${\cal S}_{\cal O}$, equals its
thermodynamic entropy, $S$. Similarly suppose a system consists of
weakly interacting subsystems, so that each subsystem can be viewed as
an isolated system in its own right. Suppose that ${\cal O}$ is
comprised by a collection of observables, ${\cal O}_i$, for each
subsytem, and suppose, in addition, that each subsystem (viewed as an
isolated system) is in thermal equilibrium -- although the entire
system need not be in thermal equilibrium.  Then ${\cal S}_{\cal O}$
will equal the sum of the thermodynamic entropies, $S_i$, of the
subsystems
\begin{equation}
{\cal S}_{\cal O} = \sum_i S_i .
\label{sumS}
\end{equation}

We now are in a position to explain the origin of the laws of
thermodynamics. As argued above, if ${\cal S}_{\cal O}$ is less than
its maximum possible value (for the given value of $E$ and the other
state parameters), we should observe it to increase until ``thermal
equilibrium'' is reached. In particular, in the case where the total
system consists of subsytems and these subsystems are individually in
thermal equilibrium at both the beginning and the end of some process
(but not necessarily at the intermediate stages), we should have
\begin{equation}
\sum_i (S_i)_1 \geq \sum_i (S_i)_0 ,
\label{2nd}
\end{equation}
where $(S_i)_0$ and $(S_i)_1$ denote, respectively, the initial and
final thermodynamic entropies of the $i$th subsystem. This accounts
for the second law of thermodynamics, eq.(\ref{t2}).

It should be noted that the time asymmetry present in this formulation
of the second law arises from the implicit assumption that, commonly,
${\cal S}_{\cal O}$ is initially below its maximum possible
value. (Otherwise, a more relevant formulation of the second law would
merely state that only very rarely and/or briefly would we expect to
observe ${\cal S}_{\cal O}$ to fluctuate below its maximum possible
value.) The fact that we do commonly observe systems with ${\cal
S}_{\cal O}$ below its maximum possible value shows that the present
state of our universe is very ``special''.

The zeroth law of thermodynamics is an immediate consequence of the
fact that if the subsystems appearing in eq.(\ref{sumS}) are at
different temperatures, then ${\cal S}_{\cal O}$ can be increased by
transferring energy from a subsystem of high temperature to a
subsystem of low temperature. (This fact follows immediately from the
definition, (\ref{temp}), of $T$.) Thus, for a thermal equilibrium
state -- where, by definition, ${\cal S}_{\cal O}$ achieves its
maximum value -- it is necessary that $T$ be uniform.

Finally, since $S$ is a function on thermodynamic state space, its
gradient can be written as
\begin{equation}
dS = \frac{1}{T} dE + \sum_j X_j d\alpha_j
\label{gradS}
\end{equation}
where $\alpha_j$ denotes the state parameters other than $E$, and $X_j
\equiv \partial S/ \partial \alpha_j$ (where $E$ and the state
parameters other than $\alpha_j$ are held fixed in taking this partial
derivative). Using Liouville's theorem, one can argue that $S$ should
be constant when sufficiently slow changes are made to parameters
appearing in the Hamiltonian. This fact gives $TX_j$ the
interpretation of being the ``generalized force'' conjugate to
$\alpha_j$ (at least for the case where $\alpha_j$ is a parameter
appearing in the Hamiltonian), and it gives $TX_j d\alpha_j$ the
interpretation of being a ``work term''. This accounts for the first
law of thermodynamics, eq.(\ref{t1}).

Thus far, our discussion has been restricted to the case of a
classical dynamical system. However, as discussed in more detail in
\cite{w1}, a completely parallel analysis can be for a quantum
system. In this analysis, the classical coarse-grained observable
${\cal O}$ on phase space is replaced by a self-adjoint operator
$\hat{\cal O}$ acting on the Hilbert space of states with energy
between $E$ and $E + \Delta E$. The spectral decomposition of
$\hat{\cal O}$ takes the form
\begin{equation}
\hat{\cal O} = \sum \lambda_m \hat{P}_m
\label{O}
\end{equation}
where the $\{\hat{P}_m\}$ are a family of orthogonal projection
operators, and it is assumed -- in correspondence with the assumptions
made about coarse-graining in the classical case -- that the
degeneracy subspaces of $\hat{\cal O}$ are large. The statistical
entropy, $\hat{\cal S}_{\hat{\cal O}}$ is then defined to be the
quantum observable
\begin{equation}
\hat{\cal S}_{\hat{\cal O}} = \sum \ln (d_m) \hat{P}_m
\label{qcalS}
\end{equation}
where $d_m$ is the dimension of the $m$th degeneracy subspace of
$\lambda_m$, i.e.,
\begin{equation}
d_m = {\rm tr} (\hat{P}_m).
\label{qcalR}
\end{equation}
Again, it is assumed that the maximum value of $d_m$ is essentially
the dimension of the entire Hilbert space of states of energy between
$E$ and $E + \Delta E$.

The corresponding definition of the thermodynamic entropy, $S$, of a
quantum system is
\begin{equation}
S = \ln n
\label{qS}
\end{equation}
where $n$ denotes the dimension of the Hilbert space of states between
$E$ and $E + \Delta E$, i.e., $n$ is proportional to the density of
quantum states per unit energy. Again, $S$ is not a quantum
observable, but rather a function on thermodynamic state
space. Arguments for the zeroth, first, and second laws of
thermodynamics then can be made in parallel with the classical case.

It should be noted that, thus far, I have made no mention of the third
law of thermodynamics. In fact, there are two completely independent
statements which are referred to as the ``third law''. The first
statement consists of the rather vague claim that it is physically
impossible to achieve $T = 0$ for a (sub)system. To the extent that it
is true, I would view this claim as essentially a consequence of the
second law, since it always is highly entropically favorable to take
energy away from a subsystem at finite temperature and add that energy
to a subsystem whose temperature is very nearly $0$. The second
statement, usually referred to as ``Nernst's theorem'', consists of
the claim that $S \rightarrow 0$ as $T \rightarrow 0$. This claim is
blatantly false in classical physics -- it fails even for a classical
ideal gas -- but it holds for many quantum systems (in particular, for
boson and fermion ideal gases). Clearly, the ``Nernst theorem'' is
actually a claim about the behavior of the density of states, $n(E)$,
as the total energy of the system goes to its minimum possible
value. Indeed, more precisely, as explained in section 9.4 of
\cite{huang}, it should be viewed as a statement about the
extrapolation to minimum energy of the continuum approximation to
$n(E)$. Elsewhere, I shall give some examples of quantum systems which
violate the ``Nernst theorem'' \cite{w2} (see also Section IV
below). Thus, while the ``Nernst theorem'' holds empirically for
systems studied in the laboratory, I do not view it as a fundamental
aspect of thermodynamics. In particular, I do not feel that the well
known failure of the analog of the ``Nernst theorem'' to hold for
black hole mechanics -- where there exist black holes of finite area
with $\kappa = 0$ -- should be viewed as indicative of any breakdown
of the relationship between thermodynamics and black hole physics.

The above discussion explains the nature and origin of the laws of
thermodynamics for ``ordinary'' classical and quantum
systems. However, as I shall now briefly describe, when general
relativity is taken into account, a number of new issues and puzzles
arise.

In the first place, it should be noted that general relativity is a
field theory and, as such, it ascribes infinitely many degrees of
freedom to the spacetime metric/gravitational field. If these degrees
of freedom are treated classically, no sensible thermodynamics should
be possible. Indeed, this situation also arises for the
electromagnetic field, where a treatment of the statistical physics of
a classical electromagnetic field in a box yields the Rayleigh-Jeans
distribution and its associated ``ultraviolet catastrophe''. As is
well known, this difficulty is cured by treating the electromagnetic
field as a quantum field. I see no reason not to believe that similar
difficulties and cures will occur for the gravitational
field. However, it should be emphasized at the outset that one should
not expect any thermodynamic laws to arise from a statistical physics
treatment of classical general relativity; a quantum treatment of the
degrees of freedom of the gravitational field should be essential.

A much more perplexing issue arises from the fact that, as emphasized
above, the arguments for the validity of thermodynamics for ``ordinary
systems'' are based upon the presence of a well defined notion of
``time translations'', which are symmetries of the dynamics. Such a
structure is present when one considers dynamics on a background
spacetime whose metric possesses a suitable one-parameter group of
isometries, and when the Hamiltonian is invariant under these
isometries. However, such a structure is absent in general relativity,
where no background metric is present. Furthermore, when the degrees
of freedom of the gravitational field are excited, one would not
expect the dynamical spacetime metric to possess a time translation
symmetry. The absence of any ``rigid'' time translation structure in
general relativity can be viewed as being responsible for making
notions like the ``energy density of the gravitational field'' ill
defined in general relativity. Notions like the ``entropy density of
the gravitational field'' are not likely to fare any better. It may
still be possible to use structures like asymptotic time translations
to define the notion of the total entropy of an (asymptotically flat)
isolated system. (As is well known, total energy can be defined for
such systems.) However, for a closed universe, it seems unlikely that
any meaningful notion will exist for the ``total entropy of the
universe'' (including gravitational entropy). If so, it is far from
clear how the second law of thermodynamics is to be formulated for a
closed universe. This issue appears worthy of further exploration.

Another important issue that arises in the context of general
relativity involves ergodic behavior. As discussed above, ordinary
thermodynamics is predicated on the assumption that generic dynamical
orbits ``sample'' the entire energy shell, spending ``equal times in
equal volumes''. However, gross violations of such ergodic behavior
occur in classical general relativity on account of the irreversible
tendency for gravitational collapse to produce singularities -- from
which one cannot then evolve back to ``uncollapsed''
states. Interestingly, however, there are strong hints that ergodic
behavior could be restored in quantum gravity. In particular, the
quantum phenomenon of black hole evaporation (see Section IV below)
provides a means of evolving from a collapsed state back to an
uncollapsed configuration.

Finally, as noted above, the fact that we commonly observe the
increase of entropy shows that the present state of the universe is
very ``special''. As has been emphasized by Penrose \cite{p}, the
``specialness'' of the present state of the universe traces back,
ultimately, to extremely special initial conditions for the universe
at the ``big bang''. This specialness of the initial state of the
universe should have some explanation in a complete, fundamental
theory. However, at present, it remains a matter of speculation as to
what this explanation might be.

The above comments already give a clear indication that there are deep
and fundamental issues lying at the interface of gravitation and
thermodynamics. As we shall see in the next two sections, the theory
of black holes gives rise to very significant further relationships
between gravitation and thermodynamics.

\section{The nature of the laws of classical black hole mechanics}

As previously indicated in the Introduction, it appears overwhelmingly
likely that the laws of classical black hole mechanics must arise, in
a fundamental quantum theory of gravity, as the classical limit of the
laws of thermodynamics applied to a system comprised by a black
hole. However, as we shall see in this section, the present
derivations of the laws of classical black hole mechanics could hardly
look more different from the arguments for the corresponding laws of
thermodynamics, as given in the previous section. Nevertheless, as I
shall emphasize here, the derivations of the laws of black hole
mechanics appear to share at least one important feature of the
thermodynamic arguments: There appears to be a ``universality'' to the
laws of black hole mechanics in that the basic form of the laws
appears to be independent of the details of the precise Lagrangian or
Hamiltonian of the underlying theory of gravity -- in a manner
analogous to the ``universality'' of the form of the laws of ordinary
thermodynamics.

In this section, we will consider theories of gravity which are much
more general than general relativity, but we shall restrict attention
to geometric theories, wherein spacetime is represented by a pair $(M,
g_{ab})$, where $M$ is a manifold and $g_{ab}$ is a metric of
Lorentzian signature. Other matter fields also may be present on
spacetime. For definiteness, I will assume that $M$ is 4-dimensional,
but all results below generalize straightforwardly to any dimension $n
\geq 2$. For our discussion of the first and second laws of black hole
mechanics, it will be assumed, in addition, that the field equations
of the theory have been obtained from a diffeomorphism covariant
Lagrangian.

In physical terms, a black hole in a spacetime, $(M, g_{ab})$, is a
region where gravity is so strong that nothing can escape. In order to
make this notion precise, one must have in mind a region of spacetime
to which one can contemplate escaping. For an asymptotically flat
spacetime (representing an isolated system), the asymptotic portion of
the spacetime ``near infinity'' is such a region. The {\em black hole}
region, ${\cal B}$, of an asymptotically flat spacetime, $(M,
g_{ab})$, is defined as
\begin{equation}
{\cal B} \equiv M - I^-({\cal I}^+) ,
\label{bh}
\end{equation}
where ${\cal I}^+$ denotes future null infinity and $I^-$ denotes the
chronological past.  The {\em event horizon}, ${\cal H}$, of a black
hole is defined to be the boundary of ${\cal B}$.

If an asymptotically flat spacetime $(M, g_{ab})$ contains a black
hole ${\cal B}$, then ${\cal B}$ is said to be {\em stationary} if
there exists a one-parameter group of isometries on $(M, g_{ab})$
generated by a Killing field $t^a$ which is unit timelike at
infinity. The black hole is said to be {\em static} if it is
stationary and if, in addition, $t^a$ is hypersurface orthogonal -- in
which case there exists a discrete ``time reflection'' isometry about
any of the orthogonal hypersurfaces. The black hole is said to be {\em
axisymmetric} if there exists a one parameter group of isometries
which correspond to rotations at infinity. A stationary, axisymmetric
black hole is said to possess the ``$t - \phi$ orthogonality
property'' if the 2-planes spanned by $t^a$ and the rotational Killing
field $\phi^a$ are orthogonal to a family of 2-dimensional
surfaces. In this case, there exists a discrete ``$t - \phi$''
reflection isometry about any of these orthogonal 2-dimensional
surfaces.

For a black hole which is static or is stationary-axisymmetric with
the $t - \phi$ orthogonality property, it can be shown \cite{c} that
there exists a Killing field $\xi^a$ of the form
\begin{equation}
\xi^a = t^a + \Omega \phi^a
\label{xi}
\end{equation}
which is normal to the event horizon, ${\cal H}$. The constant
$\Omega$ defined by eq.(\ref{xi}) is called the {\em angular velocity
of the horizon}. (For a static black hole, we have $\Omega = 0$.) A
null surface whose null generators coincide with the orbits of a
one-parameter group of isometries is called a {\em Killing horizon},
so the above result states that the event horizon of any black hole
which is static or is stationary-axisymmetric with the $t - \phi$
orthogonality property must always be a Killing horizon. A stronger
result holds in general relativity, where, under some additional
assumptions, it can be shown that the event horizon of any stationary
black hole must be a Killing horizon \cite{he}. From this result, it
also follows that in general relativity, a stationary black hole must
be nonrotating (from which staticity follows \cite{sw}, \cite{cw}) or
axisymmetric (though not necessarily with the $t - \phi$ orthogonality
property).

Now, let ${\cal K}$ be any Killing horizon (not necessarily required
to be the event horizon, ${\cal H}$, of a black hole), with normal
Killing field $\xi^a$. Since $\nabla^a (\xi^b \xi_b)$ also is normal
to ${\cal K}$, these vectors must be proportional at every point on
${\cal K}$. Hence, there exists a function, $\kappa$, on ${\cal K}$,
known as the {\em surface gravity} of ${\cal K}$, which is defined by
the equation
\begin{equation}
\nabla^a (\xi^b \xi_b) = -2 \kappa \xi^a
\label{kappa}
\end{equation}
It follows immediately that $\kappa$ must be constant along each null
geodesic generator of ${\cal K}$, but, in general, $\kappa$ can vary
from generator to generator. It is not difficult to show (see, e.g.,
\cite{w3}) that
\begin{equation}
\kappa = {\rm lim} (Va)
\label{sg}
\end{equation}
where $a$ is the magnitude of the acceleration of the orbits of
$\xi^a$ in the region off of $\cal K$ where they are timelike, $V
\equiv (- \xi^a \xi_a)^{1/2}$ is the ``redshift factor'' of $\xi^a$,
and the limit as one approaches ${\cal K}$ is taken. Equation
(\ref{sg}) motivates the terminology ``surface gravity''. Note that
the surface gravity of a black hole is defined only when it is ``in
equilibrium'' (i.e., stationary), analogous to the fact that the
temperature of a (sub)system in ordinary thermodynamics is defined
only for thermal equilibrium states.

In the context of an arbitrary metric theory of gravity, the zeroth
law of black hole mechanics may now be stated as the following theorem
\cite{c}, \cite{rw2}: {\em For any black hole which is static or is
stationary-axisymmetric with the $t - \phi$ orthogonality property,
the surface gravity $\kappa$, must be constant over its event horizon
${\cal H}$.} The key ingredient in the proof of this theorem is the
identity \cite{rw2}
\begin{equation}
\xi_{[a} \nabla_{b]} \kappa = - \frac{1}{4} \epsilon_{abcd} \nabla^c \omega^d
\label{kid}
\end{equation}
which holds on an arbitrary Killing horizon, where $\omega_a \equiv
\epsilon_{abcd} \xi^b \nabla^c \xi^d$ denotes the twist of the Killing
field $\xi^a$. For a static black hole, we have $\omega_a = 0$, and
the constancy of $\kappa$ on ${\cal H}$ follows immediately. Further
arguments \cite{rw2} similarly establish the constancy of $\kappa$ for
a stationary-axisymmetric black hole with the $t - \phi$ orthogonality
property. It should be emphasized that this result is ``purely
geometrical'', and involves no use of any field equations.

A stronger version of the zeroth law holds in general
relativity. There it can be shown \cite{bch} that if Einstein's
equation holds with the matter stress-energy tensor satisfying the
dominant energy condition, then $\kappa$ must be constant on any
Killing horizon. In particular, one need not make the additional
hypothesis that the $t - \phi$ orthogonality property holds.

An important consequence of the zeroth law is that if $\kappa \neq 0$,
then in the ``maximally extended'' spacetime representing the black
hole, the event horizon, ${\cal H}$, comprises a branch of a
``bifurcate Killing horizon''. (A precise statement and proof of this
result can be found in \cite{rw2}. Here, a {\em bifurcate Killing
horizon} is comprised by two Killing horizons, ${\cal H}_A$ and ${\cal
H}_B$, which intersect on a spacelike 2-surface, $\cal C$, known as
the {\em bifurcation surface}.) As stated above, the event horizon of
any black hole which is static or is stationary-axisymmetric with the
$t - \phi$ orthogonality property, necessarily is a Killing horizon
and necessarily satisfies the zeroth law. Thus, the study of such
black holes divides into two cases: ``degenerate'' black holes (for
which, by definition, $\kappa = 0$), and black holes with bifurcate
horizons. Again, this result is ``purely geometrical'' -- involving no
use of any field equations -- and, thus, it holds in any metric theory
of gravity.

We turn, now, to the consideration of first law of black hole
mechanics. For this
analysis, it will be assumed that the field equations of the theory
arise from a diffeomorphism covariant Lagrangian 4-form, ${\bf L}$, of
the general structure
\begin{equation}
{\bf L} = {\bf L} \left( g_{ab}; R_{abcd}, \nabla_a R_{bcde},
...;\psi, \nabla_a \psi, ...\right)
\label{lag}
\end{equation}
where $\nabla_a$ denotes the derivative operator associated with
$g_{ab}$, $R_{abcd}$ denotes the Riemann curvature tensor of $g_{ab}$,
and $\psi$ denotes the collection of all matter fields of the theory
(with indices surpressed). An arbitrary (but finite) number of
derivatives of $R_{abcd}$ and $\psi$ are permitted to appear in ${\bf
L}$. Here and below we use boldface letters to denote differential
forms and we will surpress their indices. We also shall denote the
complete collection of dynamical fields, $(g_{ab}, \psi)$ by $\phi$
(thereby surpressing the indices of $g_{ab}$ as well). Our treatment
will follow closely that given in \cite{iw}; much of the mathematical
machinery we shall use also has been extensively employed in analyses
of symmetries and conservation laws of Lagrangian systems (see, e.g.,
\cite{ff} and references cited therein).

The Euler Lagrange equations of motion, ${\bf E} = 0$, are obtained by
writing the variation of the Lagrangian in the form
\begin{equation}
\delta {\bf L} = {\bf E}(\phi) \delta \phi + d {\mbox{\boldmath
$\theta$}}(\phi, \delta \phi) .
\label{dL}
\end{equation}
where no derivatives of $\delta \phi$ appear in the first term on the
right side. Usually, the manipulations yielding eq.(\ref{dL}) are
performed under an integral sign, in which case the second term on the
right side becomes a ``boundary term'', which normally is
discarded. In our case, however, our interest will be in the
mathematical structure to the theory provided by ${\mbox{\boldmath
$\theta$}}$. Indeed, the precise form of ${\mbox{\boldmath $\theta$}}$
and the auxilliary structures derived from it will play a crucial role
in our analysis, whereas the precise form of the field equations,
${\bf E} = 0$, will not be of interest here.

The sympletic current 3-form on spacetime, ${\mbox{\boldmath
$\omega$}}$ -- which is a local function of a field configuration,
$\phi$, and two linearized perturbations, $\delta_1 \phi$ and
$\delta_2 \phi$ off of $\phi$ -- is obtained by taking an
antisymmetrized variation of ${\mbox{\boldmath $\theta$}}$
\begin{equation}
{\mbox{\boldmath $\omega$}} (\phi, \delta_1 \phi, \delta_2 \phi) =
\delta_2 {\mbox{\boldmath $\theta$}} (\phi,\delta_1
\phi)-\delta_1{\mbox{\boldmath $\theta$}} (\phi,\delta_2 \phi)
\label{omega}
\end{equation}
The (pre-)symplectic form, $\Omega$ -- which is a map taking field
configurations, $\phi$, together with a pairs of linearized
perturbations off of $\phi$, into the real numbers -- is obtained by
integrating ${\mbox{\boldmath $\omega$}}$ over a Cauchy surface,
$\Sigma$
\begin{equation}
\Omega (\phi, \delta_1 \phi, \delta_2 \phi) = \int_\Sigma
{\mbox{\boldmath $\omega$}}
\label{Omega}
\end{equation}
(This integral is independent of choice of Cauchy surface when
$\delta_1 \phi$ and $\delta_2 \phi$ satisfy the linearized field
equations.) The (pre-)symplectic form, $\Omega$, provides the
structure needed to define the phase space of the theory \cite{lw}. It
also provides the structure needed to define the notion of a
Hamiltonian, $H$, conjugate to an arbitrary vector field, $\eta^a$, on
spacetime\footnote{As discussed in \cite{lw}, it will, in general, be
necessary to choose $\eta^a$ to be ``field dependent'', in order that
it ``project'' to phase space.}: $H$ is a function on phase space
satisfying the property that about any solution, $\phi$, the variation
of $H$ satisfies
\begin{equation}
\delta H = \Omega(\phi, \delta \phi, {\cal L}_\eta \phi) ,
\label{H}
\end{equation}
where ${\cal L}_\eta$ denotes the Lie derivative with respect to the
vector field $\eta^a$.  (Equation (\ref{H}) can be put in the more
familiar form of Hamilton's equations of motion by solving it for
${\cal L}_\eta \phi$, thus expressing the ``time derivative'' of
$\phi$ in terms of functional derivatives of $H$.)

On account of the diffeomorphism covariance of ${\bf L}$, the
infinitesimal diffeomorphism generated by an arbitrary vector field,
$\eta^a$, is a local symmetry of the theory. Hence, there is an
associated, conserved {\em Noether current} 3-form, ${\bf j}$, defined
by
\begin{equation}
{\bf j} = {\mbox{\boldmath $\theta$}} (\phi, {\cal L}_\eta \phi) -
\eta \cdot {\bf L}
\label{j}
\end{equation}
where the ``$\cdot$'' denotes the contraction of the vector field
$\eta^a$ into the first index of the differential form ${\bf L}$. One
can show \cite{iw2} that ${\bf j}$ always can be written in the form
\begin{equation}
{\bf j} = d {\bf Q} + \eta^a {\bf {C}}_a ,
\label{Q}
\end{equation}
where ${\bf {C}}_a = 0$ when the equations of motion hold, i.e., ${\bf
{C}}_a$ corresponds to ``constraints'' of the theory. Equation
(\ref{Q}) defines the {\em Noether charge} 2-form $\bf Q$, which is
unique up to
\begin{equation}
{\bf Q} \rightarrow {\bf Q} + \eta \cdot {\bf X}(\phi) + {\bf Y}
(\phi, {\cal L}_\eta \phi) + d {\bf Z}(\phi, \eta) .
\label{Qamb}
\end{equation}
where $\bf X$, $\bf Y$, and $\bf Z$ are arbitrary forms which are
locally constructed from the fields appearing in their arguments (and
with $\bf Y$ being linear in ${\cal L}_\eta \phi$ and $\bf Z$ being
linear in $\eta$).  Here the term $\eta \cdot {\bf X}$ arises from the
ambiguity ${\bf L} \rightarrow {\bf L} + d{\bf X}$ in the choice of
Lagrangian, the term ${\bf Y} (\phi, {\cal L}_\eta \phi)$ arises from the
ambiguity ${\mbox{\boldmath $\theta$}} \rightarrow {\mbox{\boldmath
$\theta$}} + d{\bf Y}$ in eq.(\ref{dL}), and the term $d{\bf Z}$
arises directly from eq.(\ref{Q}).

The first law of black hole mechanics is a direct consequence of the
variational identity
\begin{equation}
\delta {\bf j} = {\mbox{\boldmath $\omega$}} (\phi, \delta \phi, {\cal
L}_\eta \phi) + d(\eta \cdot {\mbox{\boldmath $\theta$}}) ,
\label{dj}
\end{equation}
which follows directly from eqs.(\ref{dL}), (\ref{omega}) and
(\ref{j}) above. One immediate consequence of this identity, together
with eq.(\ref{H}), is that if a Hamiltonian, $H$, conjugate to
$\eta^a$ exists, it must satisfy
\begin{equation}
\delta H = \int_\Sigma [\delta {\bf j} - d(\eta \cdot {\mbox{\boldmath
$\theta$}})]
\label{dH2}
\end{equation}
From eq.(\ref{Q}) it follows, in addition, that ``on shell'' -- i.e.,
when the equations of motion hold and, hence, ${\bf {C}}_a = 0$ -- we
have
\begin{equation}
\delta H = \int_\Sigma d[\delta {\bf Q} - \eta \cdot {\mbox{\boldmath
$\theta$}}] ,
\label{dH3}
\end{equation}
so, ``on shell'', $H$ is given purely by ``surface terms''. In the
case of an asymptotically flat spacetime, the surface term,
$H_\infty$, arising from infinity has the interpretation of being the
total ``canonical energy'' (conjugate to $\eta^a$) of the spacetime.

Now, let $\phi$ be any solution to the field equations ${\bf E} = 0$
with a Killing field $\xi^a$, and let $\delta \phi$ be any solution to
the linearized field equations off $\phi$ (not necessarily satisfying
${\cal L}_\xi \delta \phi = 0$). It follows immediately from
eq.(\ref{dj}) (with $\eta^a = \xi^a$) together with the variation of
eq.(\ref{Q}) that
\begin{equation}
d[\delta {\bf Q} - \xi \cdot {\mbox{\boldmath $\theta$}}] = 0
\label{ddQ}
\end{equation}
We apply this equation to a spacetime containing a black hole with
bifurcate Killing horizon, with $\xi^a$ taken to be the Killing field
(\ref{xi}) normal to the horizon, ${\cal H}$. (As mentioned above, the
assumption of a bifurcate Killing horizon involves no loss of
generality \cite{rw2} if the zeroth law holds and $\kappa \neq 0$.)
We integrate this equation over a hypersurface, $\Sigma$, which
extends from the bifurcation surface, $\cal C$, of the black hole to
infinity. The result is
\begin{equation}
\delta H_\infty = \delta \int_{\cal C} {\bf Q}
\label{dH4}
\end{equation}
where the fact that $\xi^a = 0$ on $\cal C$ has been used.

We now evaluate the surface terms appearing on each side of
eq.(\ref{dH4}). As noted above, $H_\infty$ has the interpretation of
being the canonical energy conjugate to $\xi^a$. For $\xi^a$ of the
form (\ref{xi}), we have (see \cite{sw})
\begin{equation}
\delta H_\infty = \delta M - \Omega \delta J + ...
\label{dH5}
\end{equation}
where the ``$...$'' denotes possible additional contributions from
long range matter fields. On the other hand, it is possible to
explicitly compute $\bf Q$, and thereby show that \cite{iw}
\begin{equation}
\delta \int_{\cal C} {\bf Q} = \frac{\kappa}{2 \pi} \delta S_{\rm bh} ,
\label{dQ}
\end{equation}
where
\begin{equation}
S_{\rm bh} \equiv -2 \pi \int_{\cal C} \frac{\delta L}{\delta
R_{abcd}} n_{ab} n_{cd}
\label{Sbh} .
\end{equation}
Here $n_{ab}$ is the binormal to $\cal C$ (normalized so that $n_{ab}
n^{ab} = -2$), $L$ is the Lagrangian (now viewed as a scalar density
rather than a 4-form), and the functional derivative is taken by
formally viewing the Riemann tensor as a field which is independent of
the metric in eq.(\ref{lag}). Combining eqs.(\ref{dH4}), (\ref{dH5}),
and (\ref{dQ}), we obtain
\begin{equation}
\delta M = \frac{\kappa}{2 \pi} \delta S_{\rm bh} + \Omega \delta J + ...,
\label{first}
\end{equation}
which is the desired first law of black hole mechanics. Indeed, this
result is actually stronger than the form of the first law stated in
the Introduction, since eq.(\ref{first}) holds for non-stationary
perturbations of the black hole, not merely for perturbations to other
stationary black hole states.

For the case of vacuum general relativity, where $L = R \sqrt{-g}$, a
simple calculation yields
\begin{equation}
S_{\rm bh} = A/4 .
\label{Sbh2}
\end{equation}
However, if one considers theories with non-minimally-coupled matter
or ``higher derivative'' theories of gravity, additional curvature
contributions will appear in the formula for $S_{\rm
bh}$. Nevertheless, as eq. (\ref{Sbh}) explicitly shows, in all cases,
$S_{\rm bh}$ is given by an integral of a ``local, geometrical
expression'' over the black hole horizon.

The above analysis also contains a strong hint that the second law of
black hole mechanics may hold in a wide class of theories. Consider a
stationary black hole with bifurcate Killing horizon, but now let us
normalize the Killing field, $\xi^a$, normal to the horizon by the
local condition that $\nabla_a \xi_b = n_{ab}$ on $\cal C$ (or,
equivalently, $\nabla_a \xi_b \nabla^a \xi^b = - 2$ on $\cal H$),
rather than by the asymptotic behavior (\ref{xi}) of $\xi^a$ at
infinity. With this new normalization, eq.(\ref{dQ}) (with the
$\delta$'s removed) becomes
\begin{equation}
S_{\rm bh} = 2 \pi \int_{\cal C} {\bf Q}[\xi^a] .
\label{Sbh3}
\end{equation}
It is easy to show that for a stationary black hole, this equation
continues to hold when $\cal C$ is replaced by an arbitrary
cross-section, $\sigma$, of $\cal H$.

Now, consider a process in which an initially stationary black hole
evolves through a non-stationary era, and then ``settles down'' to
another stationary final state. Let $\xi^a$ be any vector field which
coincides with the Killing field normal to $\cal H$ (with the above,
new normalization) in the two stationary regimes. Then, by
eqs.(\ref{Sbh3}) and (\ref{Q}) (together with ${\bf {C}}_a = 0$), we
obtain
\begin{eqnarray}
\Delta S_{\rm bh} & = & 2 \pi \int_{\sigma_1} {\bf Q}[\xi^a] - 2 \pi
\int_{\sigma_0} {\bf Q}[\xi^a] \nonumber\\ & = & 2 \pi \int_{\cal H}
{\bf j}[\xi^a] ,
\label{DS}
\end{eqnarray}
where ${\sigma_0}$ and ${\sigma_1}$ are, respectively, cross-sections
of $\cal H$ in the initial and final stationary regimes. Equation
(\ref{DS}) states that the change in black hole entropy is
proportional the net flux of Noether current (conjugate to $\xi^a$)
through $\cal H$. Now, in many circumstances, the Noether current
conjugate to a suitable time translation can be interpreted as the
4-density of energy-momentum. Thus, eq.(\ref{DS}) suggests that the
second law of black hole mechanics, $\Delta S_{\rm bh} \geq 0$, may
hold in all theories which have suitable positive energy properties
(such as, perhaps, a positive ``Bondi energy flux'' at null
infinity). However, I have not, as yet obtained any general results
along these lines. Nevertheless, it has long been known that the
second law holds in general relativity \cite{h}, provided that the
matter present in spacetime satisfies the following positive energy
property (known as the ``null energy condition''): for any null vector
$k^a$, the matter stress-energy tensor, $T_{ab}$, satisfies $T_{ab}
k^a k^b \geq 0$.

\section{Quantum black hole thermodynamics}

In the previous section, we used a purely classical treatment of
gravity and matter fields to derive analogs of the laws of
thermodynamics for black holes. However, as already noted in the
Introduction, in classical physics, these laws of black hole mechanics
cannot correspond physically to the laws of thermodynamics. It is only
when quantum effects are taken into account that these subjects appear
to merge.

The key result establishing a physical connection between the laws of
black hole mechanics and the laws of thermodynamics is, of course, the
thermal particle creation effect discovered by Hawking \cite{h2}. This
result is derived in the context of ``semiclassical gravity'', where
the effects of gravitation are still represented by a classical
spacetime $(M, g_{ab})$, but matter fields are now treated as quantum
fields propagating in this classical spacetime. In its most general
form, this result may be stated as follows (see \cite{w4} for further
discussion): Consider a black hole formed by gravitational collapse,
which ``settles down'' to a stationary final state. By the zeroth law
of black hole mechanics, the surface gravity, $\kappa$, of this
stationary black hole final state will be constant over its event
horizon. Consider a quantum field propagating in this background
spacetime, which is initially in any (non-singular) state. {\em Then,
at asymptotically late times, particles of this field will be radiated
to infinity as though the black hole were a perfect black body at the
Hawking temperature, eq. (\ref{T}).} Thus, a stationary black hole
truly is a state of thermal equilibrium, and $\kappa/2 \pi$ truly is
the physical temperature of a black hole. It should be noted that this
result relies only on the analysis of quantum fields in the region
exterior to the black hole. In particular, the details of the
gravitational field equations play no role, and the result holds in
any metric theory of gravity obeying the zeroth law.

The physical connection between the laws of black hole mechanics and
the laws of thermodynamics is further cemented by the following
considerations. If we take into account the ``back reaction'' of the
quantum field on the black hole (i.e., if the gravitational field
equations are used self-consistently, taking account of the
gravitational effects of the quantum field), then it is clear that if
energy is conserved in the full theory, an isolated black hole must
lose mass in order to compensate for the energy radiated to infinity
in the particle creation process. As a black hole thereby
``evaporates'', $S_{\rm bh}$ will decrease, in violation of the second
law of black hole mechanics. (Note that in general relativity, this
can occur because the stress-energy tensor of quantum matter does not
satisfy the null energy condition -- even for matter for which this
condition holds classically -- in violation of one of the hypotheses
of the area theorem.) On the other hand, there is a serious difficulty
with the ordinary second law of thermodynamics when black holes are
present: One can simply take some ordinary matter and drop it into a
black hole, where, classically at least, it will disappear into a
spacetime singularity. In this latter process, one loses the entropy
initially present in the matter, but no compensating gain of ordinary
entropy occurs, so the total entropy, $S$, decreases. Note, however,
that in the black hole evaporation process, although $S_{\rm bh}$
decreases, there is significant amount of ordinary entropy generated
outside the black hole due to particle creation. Similarly, when
ordinary matter (with positive energy) is dropped into a black hole,
although $S$ decreases, by the first law of black hole mechanics,
there will necessarily be an increase in $S_{\rm bh}$.

The above considerations motivated the following proposal \cite{b1},
\cite{b2}. Although the second law of black hole mechanics breaks down
when quantum processes are considered, and the ordinary second law
breaks down when black holes are present, perhaps the following law,
known as the {\em generalized second law} always holds: {\em In any
process, the total generalized entropy never decreases}
\begin{equation}
\Delta S' \geq 0 ,
\label{gsl}
\end{equation}
where the {\em generalized entropy}, $S'$, is defined by 
\begin{equation}
S' \equiv S + S_{\rm bh} .
\label{S'}
\end{equation}
A number of analyses \cite{uw}, \cite{tz}, \cite{fp}, \cite{s} have
given strong support to the generalized second law.  Although these
analyses have been carried out in the context of general relativity,
the arguments for the validity of the generalized second law should be
applicable to a general theory of gravity, provided, of course, that
the second law of black hole mechanics holds in the classical theory.

The generalized entropy (\ref{S'}) and the generalized second law
(\ref{gsl}) have obvious interpretations: Presumably, for a system
containing a black hole, $S'$ is nothing more than the ``true total
entropy'' of the complete system, and (\ref{gsl}) is then nothing more
than the ``ordinary second law'' for this system. If so, then $S_{\rm
bh}$ truly is the physical entropy of a black hole.

Although I believe that the above considerations make a compelling
case for the merger of the laws of black hole mechanics with the laws
of thermodynamics, there remain many puzzling aspects to this
merger. One such puzzle has to do with the existence of a ``thermal
atmosphere'' around a black hole. It is crucial to the arguments for
the validity of the generalized second law (see, in particular,
\cite{uw}) that near the black hole, all fields are in thermal
equilbrium with respect to the notion of time translations defined by
the horizon Killing field $\xi^a$ (see eq.(\ref{xi}) above). For an
observer following an orbit of $\xi^a$ just outside the black hole,
the locally measured temperature (of all species of matter) is
\begin{equation}
T = \frac{\kappa}{2 \pi V} \: ,
\label{ta}
\end{equation}
where $V = (-\xi^a \xi_a)^{1/2}$.  Note that, in view of eq.(\ref{sg})
above, we see that $T \rightarrow a/2 \pi$ as the black hole horizon,
$\cal H$, is approached. Thus, in this limit eq.(\ref{ta}) corresponds
to the flat spacetime Unruh effect \cite{u}.

Since $T \rightarrow \infty$ as the horizon is approached, this
``thermal atmosphere'' has enormous entropy. Indeed, if no cut-off is
introduced, the entropy of the thermal atmosphere is
divergent.\footnote{If a cut-off at the Planck scale is introduced,
the entropy of the thermal atmosphere agrees, in order of magnitude,
with the black hole entropy (\ref{Sbh2}) There have been a number of
attempts (see \cite{s} for further discussion) to attribute the
entropy of a black hole to this thermal atmosphere, or to ``quantum
hair'' outside the black hole. However, if any meaning can be given to
the notion of where the entropy of a black hole ``resides'', it seems
much more plausible to me that it resides in the ``deep interior'' of
the black hole (corresponding to the classical spacetime
singularity).}. However, inertial observers do not ``see'' this
thermal atmosphere, and would attribute the physical effects produced
by the thermal atmosphere to other causes, like radiation reaction
effects \cite{uw}. In particular, with respect to a notion of ``time
translations'' which would be naturally defined by inertial observers
who freely fall into the black hole, the entropy of quantum fields
outside of a black hole should be negligible.

Thus, it is not entirely clear what the quantity ``$S$'' appearing in
eq.(\ref{S'}) is supposed to represent for matter near, but outside
of, a black hole. Does $S$ include contributions from the thermal
atmosphere? If so, $S$ is divergent unless a cutoff is introduced --
although changes in $S$ (which is all that is needed for the
formulation of the generalized second law) could still be well defined
and finite. If not, what happens to the entropy in a box of ordinary
thermal matter as it is slowly lowered toward the black hole? By the
time it reaches its ``floating point'' \cite{uw}, its contents are
indistinguishable from the thermal atmosphere, so has its entropy
disappeared? These questions provide good illustrations of some of the
puzzles which arise when one attempts to consider thermodynamics in the
framework of general relativity, as previously discussed at the end of
Section II.

However, undoubtedly the most significant puzzle in the relationship
between black holes and thermodynamics concerns the physical origin of
the entropy, $S_{\rm bh}$, of a black hole. Can the origin of $S_{\rm
bh}$ be understood in essentially the same manner as in the
thermodynamics of conventional systems (as suggested by the apparently
perfect merger of black hole mechanics with thermodynamics), or is
there some entirely new phenomena at work here (as suggested by the
radical differences in the present derivations of the laws of black
hole mechanics and the laws of thermodynamics)? As already remarked
near the end of Section II, a classical treatment (as in Section III),
or even a semiclassical treatment (as above in this Section), cannot
be adequate to analyze this issue; undoubtedly, a fully quantum
treatment of all of the degrees of freedom of the gravitational field
will be required.

Our present understanding of quantum gravity is quite
rudimentary. Most approaches used to formulate the quantum theory of
fields propagating in a Minkowski background spacetime either are
inapplicable or run into severe difficulties when one attempts to
apply them to the formulation of a quantum theory of the spacetime
metric itself. However, several approaches have been developed which
have some appealing features and which hold out some hope of
overcoming these difficulties. The most extensively developed of these
approaches is string theory, and, very recently, some remarkable
results have been obtained in the context of string theory relevant to
understanding the origin of black hole entropy. I will very briefly
describe some of the key results here, referring the
reader to the contribution of Horowitz \cite{ho} for further details
and discussion.

In the context of string theory, one can consider a ``low energy
limit'' in which the ``massive modes'' of the string are neglected. In
this limit, string theory should reduce to a 10-dimensional
supergravity theory. If one treats this supergravity theory as a
classical theory involving a spacetime metric, $g_{ab}$, and other
classical fields, one can find solutions describing black holes. These
classical black holes can be interpreted as providing a description of
states in string theory which is applicable at ``low energies''.

On the other hand, one also can consider a ``weak coupling'' limit of
string theory, wherein the states are treated perturbatively about a
background, flat spacetime. In this limit, the dynamical degrees of
freedom of the theory are described by perturbative string states
together with certain soliton-like configurations, known as
``D-branes''. In the weak coupling limit, there is no literal notion
of a black hole, just as there is no notion of a black hole in
linearized general relativity. Nevertheless, certain weak coupling
states comprised of ``D-branes'' can be identified with certain black
hole solutions of the low energy limit of the theory by a
correspondence of their energy and charges.

Now, the ``weak coupling'' states are, in essence, ordinary quantum
dynamical degrees of freedom in a flat background spacetime. The
ordinary entropy, $S$, of these states can be computed using
eq.(\ref{qS}) of Section II. The corresponding classical black hole
states of the ``low energy limit'' of the theory have an entropy,
$S_{\rm bh}$, given by eq.(\ref{Sbh2}) of Section III. The remarkable
results referred to above are the following: For certain
classes of extremal ($\kappa = 0$) \cite{stva} and nearly extremal
\cite{nonex} black holes, the ``ordinary entropy'', (\ref{qS}), of the
weak coupling D-brane states agrees exactly with the entropy,
(\ref{Sbh2}), of the corresponding classical black hole states
occurring in the low energy limit of the theory.\footnote{Note that
since the D-brane states are, in essence, ordinary quantum systems,
they thereby provide an example of a quantum system which violates the
``Nernst theorem'' formulation of the third law of thermodynamics.}
Since these entropies have a nontrivial functional dependence on
energy and charges, it is hard to imagine that this agreement could be
the result of a random coincidence. Furthermore, for low energy scattering,
the absorption/emission coefficients (``gray body factors'')
of the corresponding D-brane and black hole configurations also agree
\cite{ms}. In particular, since the temperatures of the corresponding
D-brane and black hole configurations agree (by virtue of the equality
of entropies), it follows that the ``Hawking radiation'' from the
corresponding D-brane and black hole configurations also agree, at
least at low energies.

Since the entropy of the D-brane configurations arises entirely from
conventional ``state counting'', the above results strongly suggest
that the physical origin of black hole entropy should be essentially
the same as for conventional thermodynamic systems. However,
undoubtedly, a much more complete understanding of the nature of black
holes in quantum gravity when one is neither in a ``weak coupling''
nor ``low energy'' limit will be required before a definitive answer
to this question can be given.

As I have tried to summarize in this article, a great deal of progress
has been made in our understanding of black holes and thermodynamics,
but many unresolved issues remain. I believe that the attainment of a
deeper understanding of the nature of the relationship between black
holes and thermodynamics is our most promising route toward an
understanding of the fundamental nature of both quantum gravity {\em
and} thermodynamics.

\end{document}